\documentclass[apj]{emulateapj}
\newcommand{\asec}{^{\prime\prime}}
\newcommand{\derr}[3]{#1^{+#2}_{-#3}}

\shorttitle{Dust in the polar region of AGN}
\shortauthors{H\"onig et al.}

\journalinfo{The Astrophysical Journal, 771:87 (15pp), 2013 July 10}
\submitted{Submitted 2013 February 19; accepted 2013 April 23; published online 2013 June 20}

\begin{document}

\title{Dust in the polar region as a major contributor to the infrared emission of \\ active galactic nuclei}

\author{S.~F. H\"onig,\altaffilmark{1} M. Kishimoto,\altaffilmark{2} K.~R.~W. Tristram,\altaffilmark{2} M.~A. Prieto,\altaffilmark{3} P. Gandhi,\altaffilmark{4,5} D. Asmus,\altaffilmark{2} R. Antonucci,\altaffilmark{1} \\ L. Burtscher,\altaffilmark{6}  W.~J. Duschl,\altaffilmark{7,8} G. Weigelt\altaffilmark{2}}
\altaffiltext{1}{Department of Physics, University of California in Santa Barbara, Broida Hall, Santa Barbara, CA 93109, USA; shoenig@physics.ucsb.edu}
\altaffiltext{2}{Max-Planck-Institut f\"ur Radioastronomie, Auf dem H\"ugel 69, 53121 Bonn, Germany}
\altaffiltext{3}{Instituto de Astrof{\'i}sica de Canarias, La Laguna, Tenerife, Spain}
\altaffiltext{4}{Institute of Space and Astronautical Science (ISAS), Japan Aerospace Exploration Agency, 3-1-1 Yoshinodai, chuo-ku, Sagamihara, Kanagawa 252-5210, Japan}
\altaffiltext{5}{Department of Physics, University of Durham, South Road, Durham DH1 3LE, UK}
\altaffiltext{6}{Max-Planck-Institut f\"ur extraterrestrische Physik, Gie{\ss}enbachstraße, 85748 Garching, Germany}
\altaffiltext{7}{Institut f\"ur Theoretische Physik und Astrophysik, Christian-Albrechts-Universit\"at zu Kiel, Leibnizstr. 15, 24098, Kiel, Germany}
\altaffiltext{8}{Steward Observatory, The University of Arizona, 933 N. Cherry Ave, Tucson, AZ 85721, USA}

\begin{abstract}
Dust around active galactic nuclei (AGN) is distributed over a wide range of spatial scales and can be observed in the infrared (IR). It is generally assumed that the distribution on parsec scales forms a geometrically- and optically-thick entity in the equatorial plane around the accretion disk and broad-line region -- dubbed ``dust torus'' -- that emits the bulk of the sub-arcsecond-scale IR emission and gives rise to orientation-dependent obscuration. However, recent IR interferometry studies with unprecedented position angle and baseline coverage on these small scales in two obscured (type 2) AGN have revealed that the majority of the mid-IR emission in these objects is elongated in polar direction. These observations are difficult to reconcile with the standard interpretation that most of the parsec-scale mid-IR emission in AGN originates from the torus and challenges the justification of using simple torus models to model the broad-band IR emission. Here we report detailed interferometry observations of the unobscured (type 1) AGN in NGC~3783 that allow us to constrain the size, elongation, and direction of the mid-IR emission with high accuracy. The mid-IR emission is characterized by a strong elongation toward position angle PA $-52^\circ$, closely aligned with the polar axis (PA $-45^\circ$). We determine half-light radii along the major and minor axes at 12.5\,$\micron$ of $(20.0\pm3.0)\,\mathrm{mas}\,\times\,(6.7\pm1.0)\,\mathrm{mas}$ or $(4.23\pm0.63)\,\mathrm{pc}\,\times\,(1.42\pm0.21)\,\mathrm{pc}$, which corresponds to intrinsically-scaled sizes of $(69.4\pm10.8)\,r_\mathrm{in}\,\times\,(23.3\pm3.5)\,r_\mathrm{in}$ for the inner dust radius of $r_\mathrm{in}=0.061$\,pc as inferred from near-IR reverberation mapping. This implies an axis ratio of 3:1, with about 60$-$90\% of the $8-13\,\micron$ emission associated with the polar-elongated component. It is quite likely that the hot-dust emission as recently resolved by near-IR interferometry is misaligned with the mid-IR emitting source, which also finds a correspondence in the two distinct $3-5\,\micron$ and $20\,\micron$ bumps seen in the high-angular resolution spectral energy distribution (SED). Based on this SED, we determine covering factors for the hot and warm dust components of $C_\mathrm{hot} = 0.42^{+0.42}_{-0.21}$ and $C_\mathrm{warm} = 0.92^{+0.92}_{-0.46}$, respectively. We conclude that these observations support a scenario that the majority of the mid-IR emission in Seyfert AGN originates from a dusty wind in the polar region of the AGN.
\end{abstract}

\keywords{galaxies: active -- galaxies: Seyfert -- galaxies: individual: NGC3783 -- infrared: galaxies -- techniques: high angular resolution}

\section{Introduction}

\setcounter{footnote}{0}

The mid-infrared (mid-IR) emission in radio-quiet active galactic nuclei (AGN) is associated with thermal emission from dust at parsec-scale distances from the central supermassive black hole. Due to the presence of an optically-thick obscuring medium in the equatorial plane of the AGN (commonly dubbed the ``torus''), the current paradigm suggests that the IR-emitting dust is the same medium as the obscuring material. This is reflected by the widespread use of torus models in order to reproduce IR spectral energy distributions (SEDs).

A number of nearby Seyfert galaxies have been shown to have significant amount of mid-IR emission extended in the polar region of the AGN on large scales of 100 parsecs (pc) \citep[e.g.][]{Boc00,Pac05,Hon10a,Reu10}. More recently, mid-IR interferometry showed that in two well-studied type 2 AGN (Circinus and NGC~424) this polar elongation is also present on parsec scales \citep{Tri12,Hon12}, and that the polar region contributes a significant fraction ($\ga50$\%) to the total mid-IR AGN emission. It is currently unclear if this feature is consistent with simple torus models \citep[see discussion in][]{Hon12}. Most current radiative transfer models for AGN tori assume an ad-hoc dust distribution that flares toward larger distances, commonly with scale heights of order unity. As such the dust distribution can be considered a single entity with the majority of the dust mass located close to the equatorial plane. The brightness distribution of these models usually predicts circular mid-IR emission or elongations along the torus mid-plane \citep[e.g.][]{Gra97,Sch05,Hon06,Sch08,Hon10b,Sta12,Hey12}. However, such \textit{equatorial} torus emission is only detected in cases where the emission from the polar region can be resolved out and an additional compact component is resolved, as in the Circinus galaxy and NGC~1068 \citep{Tri07,Rab09}. Torus models that are capable of reproducing the equatorial component do not show a significant polar component \citep[e.g.][]{Hon06,Hon08}. 

We recently proposed a dusty wind scenario that may qualitatively explain the polar-extended mid-IR emission seen on parsec scales \citep{Hon12}. In this picture a mostly optically-thin dusty wind is launched from the hottest and inner region of an optically thick dusty disk. While the inner, hot part of this disk may be puffed-up and provide the obscuring properties associated with the classical torus on the sub-parsec level, the dusty wind extends in polar direction over parsecs or even tens of parsecs. Accordingly, the single-telescope mid-IR emission is primarily originating from the polar wind, and the puffed-up inner disk emits the bulk of the near-IR emission. Although (magneto-)hydrodynamically-driven disk winds have been proposed in literature as a possible mechanism to obtain geometrically-thick obscuration \citep[e.g.][]{Emm92,Kon94,Elv00}, it is unclear if such a process would also be effective to explain the dust on up to 100\,pc scales in the polar region as seen in some AGN (see previous paragraph). In \citet{Hon12} we propose radiation pressure on dust as a potential alternative mechanism, which may be supported by a possible luminosity-dependence of the polar component as reported in \citet[][ see also \citet{Kis11b}]{Kis13}. Crucial tests for this scenario are (1) that the polar-extended emission is generic and frequently detected in Seyfert galaxies, and (2) that the near-IR and mid-IR are actually distinct emission regions. Owing to the small spatial scales at which the near- and mid-IR radiation is emitted, IR interferometry is the only way to directly probe how the dust is distributed on parsec scales around an AGN.

The current interferometric results pose a potential challenge the paradigm that the mid-IR emission of AGN predominantly originates from the torus. The key to distinguish between torus and dusty wind is a detection of the elongation of the parsec-scaled mid-IR emission and its detailed characterization, for which a dense coverage of interferometric $uv$-plane (i.e. observations at many position angles and baseline lengths) is required. To date suitable data are only available for Circinus and NGC~424, two Seyfert 2 AGN. In this paper we present new interferometry observations of the type 1 AGN in NGC~3783 to constrain the position-angle dependence of the mid-IR brightness distribution. Observing a type 1 instead of a type 2 AGN has the advantage that possible obscuration/extinction effects are avoided. A small set of mid-IR interferometric observations of this source have already been analyzed \citep{Bec08,Kis09a,Kis11b}, but the $uv$-coverage did not allow for a detailed study of the shape of the mid-IR emission. Here we present a total of 41 independent datasets covering essentially the full $uv$-plane in position angle.

In Sect.~\ref{sec:obs} we describe our interferometric data set that have been taken during several observing campaigns at the Very Large Telescope Interferometer (VLTI) between 2005 and 2012. Following this, we present the results and complement the mid-IR interferometry with a high-angular resolution IR spectral energy distribution (SED) and near-IR interferometry from literature. All data have been modeled in Sect.~\ref{sec:model}, and implications are discussed in Sec.~\ref{sec:disc}. Finally, our results are summarized in Sect.~\ref{sec:summary}.

\section{Observations and data reduction}\label{sec:obs}

\begin{figure}
\begin{center}
\epsscale{1.2}
\plotone{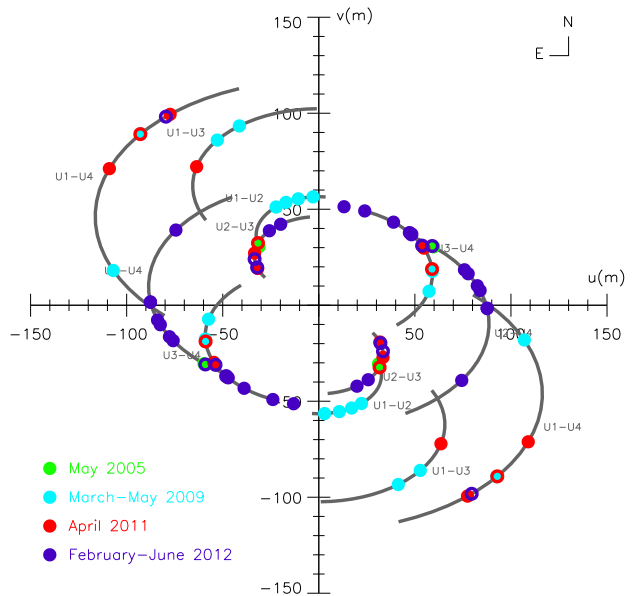}
\caption{$uv$-plane coverage (on-sky orientation) of the mid-IR interferometric observations of NGC~3783. North is up, east is left. The different colors represent the various observing runs in May 2005 (green), March and May 2009 (light blue), April 2011 (red), and February to June 2012 (violet).\label{fig:uv}}
\end{center}
\end{figure}
 
We carried out interferometric observations spread over several runs in 2005, 2009, 2011, and 2012 using the mid-IR beam combiner \textit{MIDI} at the VLTI in Chile. This instrument allows for spectro-interferometric observations in the mid-IR waveband from $8-13\,\micron$. We used all possible pairwise combinations of the four 8-m telescopes, resulting in a well-covered accessible $uv$-plane, as shown in Fig.~\ref{fig:uv}. The $uv$-plane is plotted as projected on the sky to allow for an easy comparison with position angles. While the shorter baselines of up to 60\,m cover all quadrants, the longer baselines are oriented preferentially in NE/SW direction, in line with the location of the UT telescopes at the observatory. This causes a non-circular beam shape that has to be taken into account when analyzing the data.

The data have been extracted by the standard MIA+EWS package developed for MIDI. Reduction and calibration were performed using the signal-to-noise ratio (SNR) maximization method as described in \citet{Kis11b}. Each individual fringe-track data set and a dedicated interferometric calibrator star have been reduced together to determine the required smoothing of the calibrator data and achieve the best SNR of the correlated fluxes. During all observing campaigns, we observed the same four calibrators near NGC~3783 (HD101666, HD112213, HD100407, and HD102964). Cross-calibration between these calibrators (see also Table~\ref{tab:obs}), some observed several times during one night as well as constantly over the different campaigns, showed that they remained essentially constant (limit on total flux change $\sim$5\%) over the time span from 2009$-$2012. 

When two or three science fringe track sets have been observed together (i.e. within about a 20\,min interval), they were averaged after data reduction with weights according to their individual SNR. This led to 2 independent correlated flux measurements in 2005, 10 in 2009, 10 in 2011, and 19 in 2012 as listed in Table~\ref{tab:obs}. The maximized SNR values of the individual fringes of all data sets range from 2.5 to 6.4 with a mean of 3.9. This is very typical of AGN observed with MIDI \citep[see][ and consistent results by \citet{Tri09} and \citet{Bur09} using a different reduction method]{Kis11b,Hon12} and can be understood by the mid-IR brightness of the source.

\begin{deluxetable}{c c c c c c}
\tablecaption{Observation log and data properties\label{tab:obs}}
\tablewidth{0pt}
\tablehead{
Observing Date                & telescopes       & PBL     & PA       & calibrator   & max. \\ 
(UT)                                  &                         & (m)     & (deg)   &                   & SNR  \\ 
}
\startdata
2005--05--28 03:50      &  U1--U2           & ~43.5 & ~45.5  & HD100407 & 5.8 \\ 
2005--05--31 03:11      &  U2--U4           & ~66.8 & 117.1  & HD100407 & 3.7 \\ \tableline
2009--03--13 03:26      &  U3--U4           & ~57.9 & ~97.4  & HD101666 & 3.8 \\ 
2009--03--13 04:32      &  U3--U4           & ~61.3 & 105.5  & HD101666 & 4.5 \\ 
2009--03--14 03:43      &  U1--U3           & 102.1 & ~24.0  & HD100407 & 2.9 \\ 
2009--03--14 04:41      &  U1--U3           & 100.9 & ~31.6  & HD100407 & 4.4 \\ 
2009--03--15 01:53      &  U1--U2           & ~56.5 & ~~3.1  & HD100407 & 4.4 \\ 
2009--03--15 02:48      &  U1--U2           & ~56.4 & ~11.0  & HD102964 & 4.6 \\ 
2009--03--15 03:38      &  U1--U2           & ~56.2 & ~17.6  & HD102964 & 6.4 \\ 
2009--03--15 04:23      &  U1--U2           & ~55.8 & ~23.6  & HD101666 & 5.7 \\ 
2009--05--11 03:31      &  U1--U4           & 108.3 & ~80.4  & HD112213 & 3.7 \\ 
2009--05--11 23:26      &  U1--U4           & 128.7 & ~46.1  & HD112213 & 4.2 \\ \tableline 
2011--04--14 04:09      &  U1--U3           & ~96.1 & ~41.4  & HD101666 & 3.0 \\ 
2011--04--15 03:18      &  U2--U3           & ~45.3 & ~44.4  & HD101666 & 4.7 \\ 
2011--04--15 04:18      &  U2--U3           & ~43.0 & ~50.9  & HD100407 & 3.8 \\ 
2011--04--15 05:48      &  U2--U3           & ~37.1 & ~58.9  & HD100407 & 3.9 \\ 
2011--04--16 00:26      &  U1--U4           & 125.9 & ~38.0  & HD101666 & 2.9 \\ 
2011--04--16 01:11      &  U1--U4           & 128.7 & ~46.2  & HD102964 & 2.7 \\ 
2011--04--16 02:17      &  U1--U4           & 130.0 & ~57.1  & HD102964 & 3.3 \\ 
2011--04--17 02:24      &  U3--U4           & ~61.9 & 107.8  & HD101666 & 3.4 \\ 
2011--04--17 03:35      &  U3--U4           & ~62.2 & 118.7  & HD101666 & 3.8 \\ 
2011--04--17 04:32      &  U3--U4           & ~60.5 & 128.4  & HD102964 & 4.0 \\ \tableline 
2012--02--04 05:29      &  U2--U4           & ~84.1 & ~62.4  & HD102964 & 3.1 \\ 
2012--02--05 05:08      &  U1--U4           & 126.3 & ~39.1  & HD101666 & 4.0 \\ 
2012--04--05 01:37      &  U2--U3           & ~46.6 & ~25.3  & HD101666 & 3.6 \\ 
2012--04--05 02:32      &  U2--U3           & ~46.6 & ~33.6  & HD102964 & 3.4 \\ 
2012--05--04 02:23      &  U3--U4           & ~62.2 & 118.4  & HD101666 & 3.0 \\ 
2012--05--04 03:17      &  U3--U4           & ~60.3 & 128.9  & HD102964 & 2.5 \\ 
2012--05--04 04:11      &  U3--U4           & ~58.1 & 138.1  & HD101666 & 4.1 \\ 
2012--05--06 02:12      &  U2--U4           & ~87.6 & ~89.0  & HD101666 & 4.6 \\ 
2012--05--06 02:53      &  U2--U4           & ~84.2 & ~95.3  & HD102964 & 4.8 \\ 
2012--05--06 03:33      &  U2--U4           & ~79.3 & 102.0  & HD102964 & 4.5 \\ 
2012--06--02 01:41      &  U2--U3           & ~41.1 & ~54.3  & HD101666 & 4.9 \\ 
2012--06--02 02:30      &  U2--U3           & ~37.2 & ~58.8  & HD102964 & 4.1 \\ 
2012--06--04 01:22      &  U3--U4           & ~60.4 & 129.0  & HD101666 & 3.7 \\ 
2012--06--04 03:18      &  U3--U4           & ~54.5 & 154.4  & HD101666 & 3.6 \\ 
2012--06--04 04:02      &  U3--U4           & ~52.9 & 165.8  & HD102964 & 4.1 \\ 
2012--06--05 00:15      &  U2--U4           & ~87.5 & ~89.1  & HD101666 & 2.6 \\ 
2012--06--05 01:06      &  U2--U4           & ~83.0 & ~97.1  & HD102964 & 3.0 \\ 
2012--06--05 01:45      &  U2--U4           & ~77.9 & 103.8  & HD101666 & 3.1 \\ 
2012--06--05 02:53      &  U2--U4           & ~66.4 & 117.7  & HD101666 & 3.2 \\ 
\enddata
\end{deluxetable}
 
\begin{figure}
\begin{center}
\epsscale{1.2}
\plotone{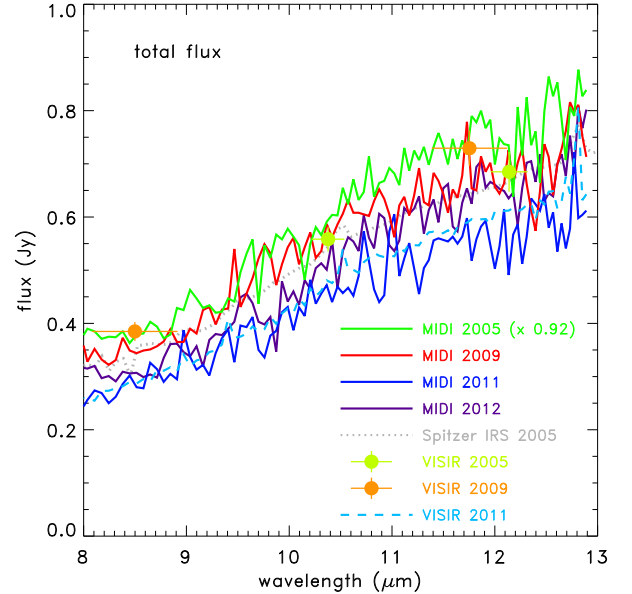}
\caption{Total $8-13\,\micron$ flux spectrum of NGC~3783 as measured with MIDI in 2005 (green solid line), 2009 (red solid line), 2011 (blue solid line), and 2012 (purple solid line). The 2005 flux has been adjusted by a factor of 0.92 to match the almost simultaneous VISIR mid-IR photometry (light-green filled circles) of higher data quality. The Spitzer spectrum of the same year is overplotted as the gray-dotted line. Also shown are VISIR fluxes from 2009 (orange filled circles) and a VISIR spectrum of 2011 (light blue dashed line) that was extracted matching the MIDI total flux aperture.}\label{fig:tflux}
\end{center}
\end{figure}

We extracted the total fluxes from all of the photometry data that have been taken together with the fringe tracks. As sky reference we used the background flux from a 5 to 7 pixel offset position to the peak of the target flux, depending on the width of the AGN emission. Calibration was done using the same calibrator star as used for the interferometric data. The resulting mid-IR spectra are shown in Fig.~\ref{fig:tflux}. For 2005, 2009, and 2011 we do have independent flux measurements of high quality to test the reliability of the MIDI fluxes: In 2005 and 2009, VLT/VISIR images were observed independently from the MIDI data. For the 2011 MIDI campaign, we took a dedicated VISIR low-resolution spectrum around the time of our interferometric observations. This spectrum has been reduced and calibrated using the dedicated pipeline described in \citet{Hon10a} and extracted from a $0\farcs75 \times 0\farcs53$ window to closely match the MIDI window of $0\farcs77 \times 0\farcs52$. 

While the 2009, 2011, and 2012 data show good consistency between the different instruments, the 2005 VISIR fluxes are about a factor of 0.92 systematically lower than the MIDI spectrum. A Spitzer IRS spectrum taken within a month of the MIDI observations shows consistency with the VISIR fluxes. Given that the 2005 MIDI spectrum is an SNR-weighted average of only 4 individual data sets, and known issues with MIDI total fluxes when only few sets are available \citep[see a similar case in][]{Hon12}, we corrected the MIDI total flux spectrum of 2005 by a factor of 0.92 to match the higher quality 2005 VISIR photometry.
 
Interestingly, the 2011 MIDI and VISIR spectra appear to be systematically lower than the 2005, 2009, and 2012 data. This may indicate mid-IR variability of the nuclear emission. \citet{Gla04} reports $K$-band variability of 0.9 magnitudes over several year time scales on fluxes uncorrected for the constant host contribution. With more than a factor of 2 variability in the near-IR, it is plausible that we may also detect variability in the mid-IR \citep{Hon11b}. We initiated an IR monitoring campaign for NGC 3783 and will discuss variability in this source in an upcoming paper. For the purpose of getting the best reference for the correlated fluxes, we will use the 2005 and 2009 MIDI total fluxes for the respective epochs, and the VISIR spectrum for our 2011 and 2012 data.

\section{Results}\label{sec:res}

\subsection{Nuclear infrared SED}\label{sec:irsed}

\begin{figure}
\begin{center}
\epsscale{1.2}
\plotone{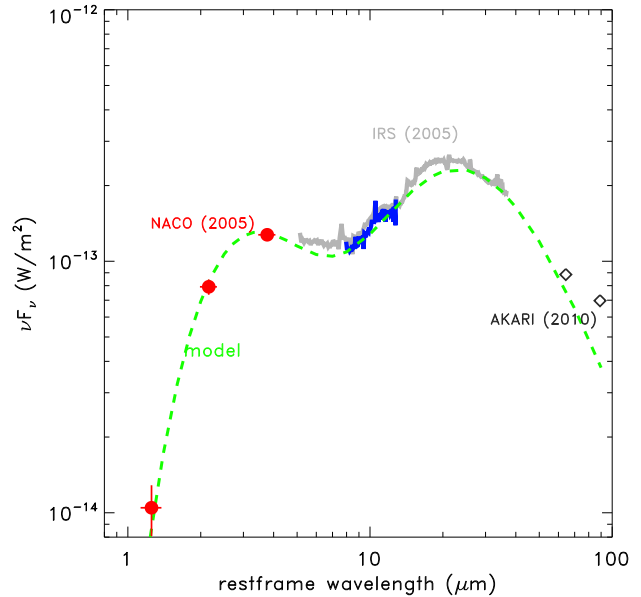}
\caption{$2-90\,\micron$ broad-band SED of NGC~3783. The gray line shows the Spitzer IRS spectrum (spatial resolution $\sim1\farcs2-10\,\asec$), while the blue line represents the VISIR spectrum from 2011 (spatial resolution $\sim0\farcs27-0\farcs4$). The red circles with error bars are nuclear photometry from \citet{Pri10} using the VLT/NACO instrument and corrected for flux contribution from the big blue bump emission based on the decomposition presented in \citet{Wei12}. The black diamonds represent far-IR photometry obtained by AKARI. Years in which the observations were taken are indicated. The SED is overplotted by our best-fit model (green-dashed lines; see Sect.~\ref{sec:model} for details).}\label{fig:irsed}
\end{center}
\end{figure}

In Fig.~\ref{fig:irsed} we show a broad-band IR SED from about $1\,\micron$ to $90\,\micron$ combining Spitzer and AKARI data at comparably low angular resolution with high-spatial resolution data obtained with 8m-telescopes at the VLT observatory in the near- and mid-IR. Although the IR emission in radio-quiet AGN is generally dominated by dust re-emission, the optical/UV ``big-blue bump'' (BBB) can reach into the near-IR \citep{Kis07,Kis08}. Therefore, we included the near-IR photometry presented by \citet{Pri10} and corrected it for BBB contamination according to the decomposition by \citet{Wei12}. The corrected photometry is shown in Fig.~\ref{fig:irsed}. 

\begin{figure}
\begin{center}
\epsscale{1.2}
\plotone{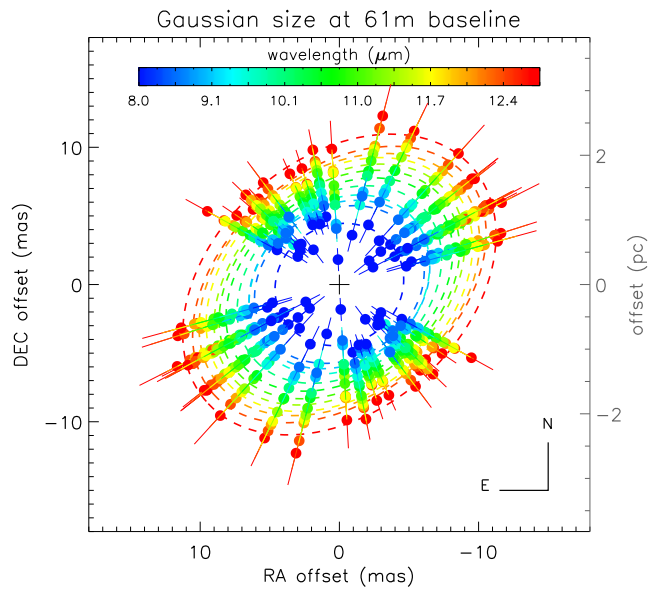}
\caption{Gaussian HWHM sizes of the mid-IR emission of NGC~3783 for different wavelengths from 8.0\,$\micron$ (blue circles) to 12.8\,$\micron$ (red circles) at a fixed baseline of 61$\pm$5\,m for a range of position angles. The colored-dashed lines are elliptical fits to the data at the corresponding wavelength bin.}\label{fig:res:allsizes}
\end{center}
\end{figure}

 Despite the differences in angular resolution, all the data connect continuously, indicating that the nuclear IR emission originates predominantly from the AGN without contamination from star formation or other host-galactic sources. This is further supported by the absence of PAH features, except for a weak 7.7\,$\micron$ line. \citet{Wu09} propose a method to use the 11.2\,$\micron$ PAH feature luminosity to estimate the star formation contribution and, in turn, obtain an AGN contribution to the mid-IR emission by comparing the PAH luminosity to the observed 12\,$\micron$ continuum luminosity. Although not present in their sample, we can use the absence of any detectable 11.2\,$\micron$ PAH feature in NGC~3783 and compare it with similar sources with comparable upper limits in \citet{Wu09}. Based on this comparison we estimate the AGN contribution to the mid-IR emission as $>$90\%. Finally, the far-IR SED is monotonically decreasing starting at $\sim20\,\micron$ toward long wavelengths, unlike starburst-contaminated sources that show much redder far-IR SEDs. Therefore, the presented SED can be considered to represent dust emission exclusively heated by the AGN.

The IR SED in Fig.~\ref{fig:irsed} shows two distinct spectral bumps. The first one peaks around $3-5\,\micron$, the second one at $\sim20\,\micron$, with a ``depression'' in the $7\,\micron$ range. Similar behavior is seen in other type 1 AGN \citep[e.g.][]{Ede86,Kis11b}, although a lack of continuous data from the near- to mid-IR often does not reveal the two peaks as clear as in the presented case of NGC~3783. This double-peaked SED suggests that the AGN-heated IR emission does not originate from a single, continuous distribution of dust.

\begin{figure*}
\begin{center}
\epsscale{0.9}
\plotone{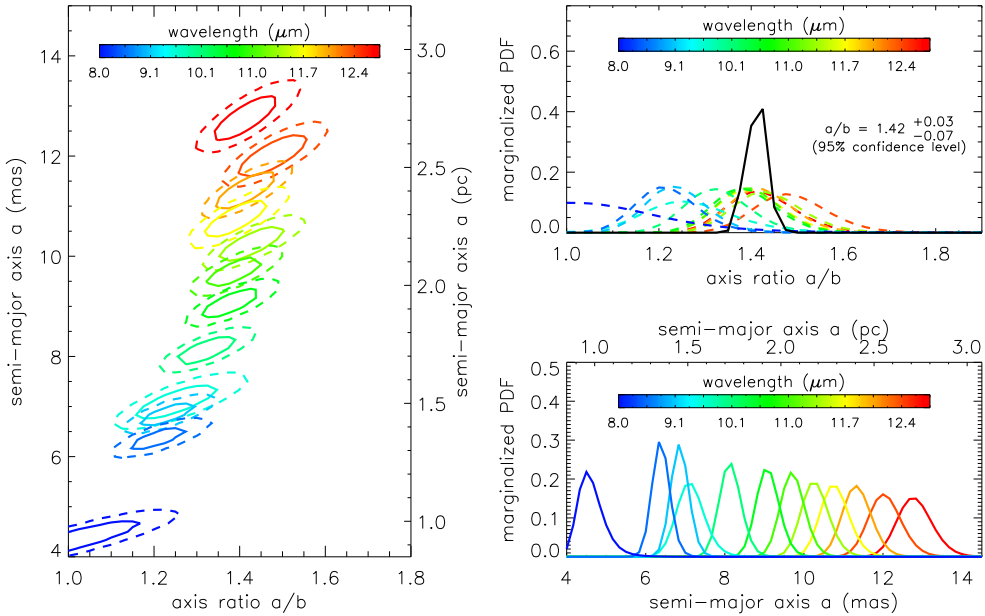}
\caption{Probability density distributions of the modeled semi-major axis $a$ and axis ratio $a/b$ for the 61\,m baseline data of NGC~3783. \textit{Left:} $a$ versus $a/b$ PDFs for different wavelengths; \textit{Upper right:} Wavelength-dependent marginalized PDFs (colored-dashed lines) and joint PDF (black-solid line) for the axis ratio $a/b$; \textit{Lower right:} Wavelength-dependent marginalized PDFs (colored-dashed lines) of the semi-major axis $a$. The color coding for the wavelengths is indicated in the color bar in all panels.} \label{fig:midi_inference}
\end{center}
\end{figure*}
 
\subsection{The size and orientation of the mid-IR emitting source in NGC~3783}\label{sec:res:ellipse}

The visibilities and correlated fluxes of all reduced 41 datasets are shown in the appendix Fig.~\ref{fig:app:intf}. These data are being analyzed in the following.

As discussed in previous papers, sizes obtained from MIDI interferometry are physically meaningful only when compared either at the same intrinsic (e.g. half-light radius) or observed (e.g. fixed baseline length) reference to account for the inhomogeneous $uv$-coverage and the unknown brightness distribution of the source \citep[for more details, please follow the discussion in][]{Kis11b,Hon12}. Owing to the lack of suitable data, it is not possible to establish a size for the half-light radius at all position angles (PAs) directly from the observations (in Sect.~\ref{sec:model} we will determine the half-light radius with the assistance of a model). Therefore, we calculate the position-angle-dependent Gaussian half width at half maximum (HWHM) sizes of the mid-IR emission source in NGC~3783 for a fixed baseline length of 61\,m. The results are shown in Fig.~\ref{fig:res:allsizes}. A two-dimensional Gaussian has been fit to all data with projected baseline length of $61\pm5$\,m. If within a position angle of $\pm5^\circ$ data at shorter and longer baseline lengths were available, we interpolated these data to 61\,m (in log-space) and used it as further input to the fit. In addition to the sizes in milliarcseconds (mas), the right ordinate axis in Fig.~\ref{fig:res:allsizes} provides a scale in pc. The dashed-colored lines are geometric fits for each wavelength bin using an elliptical Gaussian model with the semi-major axis $a$, the ratio $a/b$ of major and minor axes, and the position angle of the major axis as free parameters. 

In Fig.~\ref{fig:midi_inference} we show confidence intervals for the semi-major axis $a$ and the axis ratio $a/b$ for the 61\,m data. For each wavelength bin (color-coded with increasing wavelength from blue to red) we calculate a probability distribution function (PDF) based on the three-parameter geometric model. The left panel of Fig.~\ref{fig:midi_inference} compares the 1$\sigma$ (solid lines) and 2$\sigma$ (dashed lines) confidence intervals for $a/b$ versus $a$, i.e. marginalized over the position angle. At wavelengths $>$8.4\,$\micron$, the elongation of the mid-IR source is significant at $>3\sigma$. Moreover, a trend is obvious that the size increases with wavelength. In the lower-right panel, we plot the marginalized PDF for the semi-major axis. The peaks of the PDF for each wavelengths shifts to larger sizes from short to long wavelengths. This trend is expected for emission from centrally-heated dust, meaning that the hotter dust (emitting at shorter wavelengths) is closer to the AGN. 

\begin{deluxetable*}{c c c c c c c c}
\tablecaption{Geometric Gaussian-fit parameters of our mid-IR MIDI data and the near-IR AMBER data published in \citet{Wei12}.\label{tab:res:ellfit}}
\tablewidth{0pt}
\tablehead{
data type                     & wavelength  & \multicolumn{3}{c}{semi-major axis $a$}      &  & axis ratio $a/b$                                & position angle \\ \cline{3-5}
                            & ($\micron$) &       (mas)        &         (pc)        & ($r_\mathrm{in}$) & & & (deg)                \\ }
\startdata
MIDI PBL 61\,m   &  8.6 & $\derr{6.3}{0.3}{0.2}$ & $\derr{1.34}{0.06}{0.05}$ & $\derr{22.0}{1.0}{0.7}$ & & $\derr{1.20}{0.08}{0.05}$ & $\derr{-62}{15}{13}$ \\
                           & 10.5 & $\derr{9.0}{0.4}{0.2}$ & $\derr{1.91}{0.08}{0.04}$ & $\derr{31.3}{1.2}{0.7}$ & & $\derr{1.38}{0.07}{0.06}$ & $\derr{-50}{8}{7}$ \\
                           & 12.4 & $\derr{12.0}{0.5}{0.4}$ & $\derr{2.55}{0.10}{0.08}$ & $\derr{41.8}{1.6}{1.4}$ & & $\derr{1.48}{0.08}{0.07}$ & $\derr{-48}{6}{6}$ \\
                           & 8.0--12.8 & \multicolumn{3}{c}{$\ldots$}        & &$\derr{1.42}{0.01}{0.03}$ & $\derr{-52}{2}{3}$ \\ \tableline
AMBER                &  2.2 & $\derr{0.73}{0.35}{0.20}$ & $\derr{0.16}{0.07}{0.04}$ & $\derr{2.5}{1.2}{0.7}$ & & $\derr{1.45}{0.60}{0.36}$ & $\derr{48}{36}{77}$ \\ 
\enddata
\tablecomments{The values for the mid-IR source are based on full marginalization of the PDF, while the near-IR parameters refer to the peak in the multi-dimensional PDF. The errors indicate the 1$\sigma$ confidence interval for each parameter.}
\end{deluxetable*}
 
In the upper-right panel of Fig.~\ref{fig:midi_inference}, we show the marginalized PDF for the axis ratio $a/b$. Assuming statistical independence of each wavelength bin, we calculated a joint PDF (black-solid line) and inferred an axis ratio at 61\,m baseline length for the mid-IR emission source of NGC~3783 of $1.42^{+0.03}_{-0.07}$ at the 95\% confidence level. We caution, however, that for a given data set the errors do have a systematic component from the uncertainty of the transfer function that leads to correlations in the errors of different wavelengths. This is not included in the error estimates. As seen in both the left and upper-right panels of Fig.~\ref{fig:midi_inference}, there may be a trend of stronger elongation with increasing wavelength. Most notably the PDF of the shortest and longest wavelengths barely overlap. Similar trends have been found in other AGN on similar spatial scales \citep[NGC~1068, Circinus, and NGC424;][]{Rab09,Tri12,Hon12} and may be explained by contribution from a hot-dust component at the shorter wavelength end that has a different orientation (see Sect.~\ref{sec:nirintf}, \ref{sec:model}, and \ref{sec:disc}).

In Table~\ref{tab:res:ellfit} we list the fitted geometric parameters at 8.6$\,\micron$, 10.5$\,\micron$ and 12.4\,$\micron$ restframe wavelength to illustrate our results. The values for each parameter were obtained as the peak of the fully marginalized PDF. The errors indicate the 1$\sigma$ confidence interval. For the semi-major axis, the sizes are listed in observed units (mas), physical units (pc), and intrinsic units ($r_\mathrm{in}$). The latter ones use $r_\mathrm{in} = 0.061\,\mathrm{pc}$, which is the inner radius based on the optical luminosity of NGC~3783 and a fit to reverberation-measured inner radii of AGN \citep{Sug06,Kis09a}, approximately consistent with the $K$-band reverberation mapping of \citet{Gla92}. We also list the values of the axis ratio and position angle based on the joint inference of all wavelength bins in the range of $8.0-12.8\,\micron$. This establishes that the mid-IR emission region observed at 61\,m baseline is elongated along the major axis at position angle $-52^{+2}_{-3}\,^\circ$ with an axis ratio of $1.42^{+0.01}_{-0.03}$ ($1\sigma$ level). Although the direction of the major axis is approximately aligned with the direction where the $uv$-plane has only coverage at about 60\,m baseline length, we note that this result is mostly unaffected by the beam shape, because we made sure that the analysis has comparable resolution at all position angles. Moreover, the elongation and position angle are already apparent in the raw data (see Fig.~\ref{fig:app:intf}): Although the data at PA$>110^\circ$ comes from baseline lengths of only 60--66\,m, the visibilities are of the same level or even lower than those at 128\,m at about PA 45$^\circ$. If the object would be circular, we would expect the opposite, i.e. significantly higher visibilities at the shorter wavelengths\footnote{This is valid for the brightness distribution of NGC~3783 because we see a clear decrease of visibility with baseline length when the position angles are very similar.}.

\subsection{Comparison of the mid-IR emission region to the near-IR and AGN system axis}

\subsubsection{Near-IR interferometry}\label{sec:nirintf}

\begin{figure}
\begin{center}
\epsscale{1.2}
\plotone{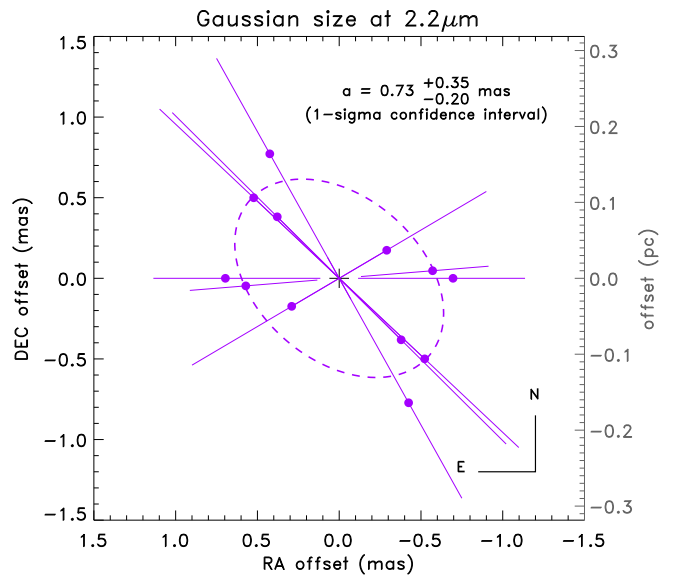}
\caption{Position angle-dependent Gaussian HWHM sizes at 2.2\,$\micron$ based on near-IR VLTI/AMBER interferometry data from \citet{Wei12} (5\% error in visibility assumed). The dashed line is the best-fit geometrical fit to the data.}\label{fig:res:amber}
\end{center}
\end{figure}
 
NGC~3783 has recently been interferometrically observed in the near-IR using the VLTI/AMBER instrument. \citet{Wei12} report a source size in the $K$-band of $0.74\pm0.23$\,mas or $0.16\pm0.05$\,pc from the two- and three-telescope interferometry. This size is about an order of magnitude smaller than the 61\,m HWHM we found in the mid-IR, qualitatively consistent with the idea that the dust is hottest closest to the AGN. The AMBER data covers a range of different position angles, and we aim for investigating if the object shows signs of ellipticity in the near-IR. We take the visibility measurements in the $K$-band from \citet{Wei12}, convert them into Gaussian HWHM, and plot the results in Fig.~\ref{fig:res:amber}. 

The large error bars of these observations ($\pm0.09$ in visibility), combined with the high visibilities, make it difficult to interpret the near-IR emission region in terms of a non-circular model. Therefore, we use the same approach as for the MIDI data and calculate the PDF for the geometric model. The parameters for the peak value in the multi-dimensional PDF are listed in Table~\ref{tab:res:ellfit}. The 1$\sigma$ confidence interval is taken from the peak of the multi-dimensional PDF along the respective parameter axis. The reason we chose this approach is that when fully marginalizing the PDFs, the inference for a position angle becomes very low and multi-peaked. This is mostly due to the lack of data between $-45^\circ$ and 20$^\circ$. Interestingly though, when marginalizing we find that the \textit{lowest} probability for the direction of the major axis is approximately at PA $-50^\circ$ to $-60^\circ$, i.e. the \textit{preferred} direction of the mid-IR extension. This may be considered as an indication that the near-IR and mid-IR are originating from distinctly different regions and dust distributions, and we want to express the need to fill the $K$-band $uv$-plane in this object.

A rather well-constrained parameter is the semi-major axis. Full marginalization provides $a=0.73^{+0.35}_{-0.20}$\,mas (1$\sigma$ confidence interval). This result is consistent with the size obtained by \citet{Wei12}. Accordingly, the $K$-band interferometry size is about a factor of 2 larger than the inferred $K$-band reverberation size \citep[based on][]{Kis07,Kis09a}, which might be explained in terms of a shallow (i.e. very extended) dust distribution, as well as a high sensitivity of the reverberation cross-correlation analysis to the fastest reacting signal \citep{Kis11a,Wei12}. An additional possibility we propose here is an inclination effect. The geometric fit provides marginal evidence that the $K$-band source is elongated. If this is indeed the case, it may be caused by inclination of the inner torus with respect to the observer. This would result in the fastest response to variability coming from the near-side along the minor axis, which according to our fit is a factor of $\derr{1.45}{0.60}{0.36}$ smaller than the semi-major axis. Further $K$-band interferometry is required, however, to formally constrain the shape of the near-IR emitting region.

\subsection{Optical polarimetry and narrow-line emission}

As discussed in \citet{Hon12}, there are at least three independent ways to establish a system axis (= polar direction) of the AGN: (1) jet-like linear radio emission in the nucleus, (2) spatial extension of the narrow-line region (NLR), and (3) optical polarimetry. However, the different methods do not always result in a consistent picture as famously evidenced in NGC~1068, where large-scale and small-scale direction indicators are misaligned. In the following we will discuss the situation in NGC~3783.

Radio maps at 3.6\,cm and 6\,cm show an unresolved point-source, so that no position angle information can be extracted \citep{Mog00,Kin00}. \citet{Smi02} report optical spectro-polarimetry of NGC~3783. The wavelength-dependent features in the percentage polarization, polarized flux, and position angle of the broad H$\alpha$ emission line are interpreted as consistent with scattering off an equatorial disk on about the same scale or slightly larger than the broad-line region (BLR) \citep{Smi02,Smi05}. As such the polarization traces the geometry of material on sub-parsec scales. The inferred polar axis from these measurements is pointed toward PA $-45^\circ$ \citep{Smi04}. 
 
Spatially-resolved observations of NGC~3783 in several optical and near-IR emission lines (e.g. [\ion{O}{3}], [\ion{Si}{6}], H$_2$) have been reported and the major axis of the extended line region is found to be between PA $-20^\circ$ to $0^\circ$ \citep[e.g.][]{Sch03,Hic09,Mue11}. Several of these lines show blue-shifted cores \citep[e.g.][]{Reu03,Hic09} and high velocities of up to 1300 km/s \citep{Kra01,Rod06,Mue11}, meaning that gas in these extended line-emitting regions is outflowing. Integral-field spectroscopy of the [\ion{Si}{6}] line \citep{Mue11} shows extended line emission over 50$-$70\,pc to the North (PA$\sim$3$^\circ$ from kinematic modeling). However, intensity maps suggest that the position angle changes toward the North/West for smaller scales \citep[see Fig.~13 in][]{Mue11}. At the $0\farcs1$/20\,pc scale, the point-spread function (PSF; FWHM $0\farcs085$) appears elongated toward PA$\sim-35^\circ$ (see the 50\% peak flux contour in the lower-right panel of their Fig.~13 and the low-velocity maps in the same figure). This is also indicated in the results of the kinematic modeling where an additional small-scale component might be necessary to explain the near-nuclear [\ion{Si}{6}] emission (F. Mueller-Sanchez, private communication).

\begin{deluxetable*}{c c c c c c c c c c}
\tablecaption{Best-fit parameters for the modeling of the IR photometry and interferometry.\label{tab:mod:fit}}
\tablewidth{0pt}
\tablehead{
model                                  & component & $\alpha$ & $\beta$ & $\phi$ & $s=a/b$ & $T_\mathrm{max}$ & $f_0$ & $R_\mathrm{in}$ & $\chi^2_r$\\
                                            &                     &        &        &  (deg)   &        &   (K)                &                          & (pc) & \\ }
\startdata
2-component& warm/mid-IR & $-0.52\pm0.24$ & $-0.28\pm0.08$ & $-56\pm1$ & $3.0\pm0.2$ & $438\pm84$ & $0.21\pm0.06$ & $0.146\pm0.063$ & 0.76 \\
 power-law                                 &  hot/near-IR   & $-7.9\pm6.3$ & $-0.5$ (fixed)  & $48$ (fixed) & $1.45$ (fixed) & $921\pm63$ & $0.09\pm0.33$ & $\ldots$           &         \\ \cline{2-10}
                                            & warm/mid-IR & $-0.41\pm0.13$ & $-0.26\pm0.06$ & $-56\pm1$ & $4.5\pm0.7$ & $414\pm50$ & $0.25\pm0.09$ & $0.092\pm0.010$ & 0.93 \\
                                            &  hot/near-IR   & $-1.2\pm0.1$ & $-0.5$ (fixed)  & $48$ (fixed) & $1.45$ (fixed) & $1400$ (fixed) & $0.08\pm0.02$ & $\ldots$           &         \\
\enddata
\tablecomments{Model parameters: $\alpha\ldots$power law index of the hot/warm surface density distribution; $\beta\ldots$temperature-gradient power law index (fixed to black-body value $\beta=-0.5$ for hot component); $\phi\ldots$position angle of the major axis in degrees from N toward E (fixed for hot component according to the geometric fit to the $K$-band interferometry); $s\ldots$ratio of major axis $a$ and minor axis $b$ (fixed for hot component according to geometric fit to $K$-band interferometry); $T_\mathrm{max}\ldots$maximum temperature of the hot/warm dust distribution; $f_0\ldots$normalized surface density, or surface emissivity, at the inner radius $R_\mathrm{in}$ for the hot/warm component; $r_\mathrm{in}\ldots$inner radius of the dust distribution;}
\end{deluxetable*}
 
The situation in NGC~3783 bears some resemblance to the changes of position angle on about the same scales seen in NGC~1068 \citep[e.g.][]{Gal04}. In the case of NGC~1068 the best inference on the parsec scale polar axis can be obtained when considering the nuclear polarization direction while the orientation of narrow-line emission on tens of parsecs scales is off by up to 45$^\circ$ \citep[e.g.][]{Mil91,Cap97,Gal04}. In NGC~3783 we find that the polarization axis, tracing the orientation on sub-pc scale, is only $\Delta$PA = $-7^\circ$ off the average mid-IR axis of our MIDI observations, and about $\Delta$PA = 87$^\circ$ off the near-IR axis, all of these directions tracing parsec-scale or even smaller structures. The orientation on tens of parsecs as inferred from the narrow-line emission intensities is off the polarization position angle by about $\Delta$PA$\sim$10$^\circ$ (smallest scales) to $48^\circ$ (larger scales; kinematic model implies opening angle of $\pm34^\circ$). The larger scale NLR axis is located somewhere in between the mid-IR and near-IR, while the polarization shows good alignment with the extended mid-IR pc-scale emission and potentially the small-scale NLR, and is about perpendicular to the near-IR.

Near- and mid-IR interferometry observations trace dust on sub-parsec to parsec scales close to the optical scattering region in this type 1 AGN. Therefore we interpret the polarization position angle in PA $-45^\circ$ as being the best reference for the direction of the polar region for our observations. Comparing this polar axis with the interferometry suggests that the mid-IR light is emitted from a region close to the polar axis while the near-IR emission originates from the AGN plane (although we note again that the inference on the position angle of the near-IR emission is not good at the moment). On larger scales this preferential axis is probably mis-aligned with the small-scale polar axis \citep[see also][]{Reu10}, indicating some kind of realignment of the material during the inflow/outflow process.

\section{Modeling the SED and interferometry of NGC~3783}\label{sec:model}

In order to use all available interferometric and photometric information simultaneously, we model the data as outlined in \citet{Hon12}. It uses a radial power law for the brightness distribution of the IR emission of the shape $S_\nu(r) = f_0 \cdot B_\nu(T(r)) \cdot \left(r/r_\mathrm{in}\right)^\alpha$, where $r$ is the distance from the AGN, $r_\mathrm{in}$ the inner radius of the dust distribution, $f_0$ is the normalized surface density (or surface emissivity) at $r_\mathrm{in}$, $\alpha$ is the surface density power law index, and $B_\nu$ the black-body radiation with temperature $T(r)$ at distance $r$. The $r$-dependent temperature $T(r)$ is parameterized as $T(r) = T_\mathrm{max} \cdot (r/r_\mathrm{in})^\beta$. The hot near-IR emission bump seen in NGC~3783 and many other type 1 AGNs is treated as a second emission component of the same form (i.e. $T_\mathrm{hot}$ as the maximum temperature, $\alpha_\mathrm{hot}$ as the surface density power law, and $f_\mathrm{0;hot}$ as the normalized surface density/emissivity of the hot component). The only difference is that we fix the temperature power law to the black-body equivalent with $\beta_\mathrm{hot} = -0.5$, which has been found a good assumption empirically \citep{Kis11b}. The geometry of the emission is modeled as elliptical with axis ratios $s$ and $s_\mathrm{hot}$ for the warm and hot component, respectively, as well as orientations $\theta$ and $\theta_\mathrm{hot}$.

We note that the use of gray-body emission ($f_0\times B_\nu(T(r))$) is a simplification. In Fig.~\ref{fig:irsed}, we can see hints of a very shallow silicate emission feature. However, in \citet{Hon10a} we analyzed the VISIR spectrum of NGC~3783 in detail and showed that the feature is extremely weak (nominally consistent with no feature or an absorption feature), which makes us confident that this assumption does not significantly influence the modeling results.

As input data we use all MIDI visibilities (12 bins in the $8-13\,\micron$ range; see Fig.~\ref{fig:app:intf}) and the MIDI total flux as listed in Table~\ref{tab:obs}. These data are complemented by the near-IR NACO data as discussed in Sect.~\ref{sec:irsed}. Since the Spitzer IRS spectrum connects smoothly to the data at higher angular resolution, we also use Spitzer fluxes at 5, 6, 7, 22, and 32\,$\micron$ to extend the SED and better sample the overall shape of both IR bumps (see Fig.~\ref{fig:irsed}).

In Table~\ref{tab:mod:fit} we list the resulting best-fit set of parameters of our model to the data. For these fits we fixed the geometry of the hot component to the position angle and axis ratio obtained by the geometric fit to the $K$-band interferometry (see Sect.~\ref{sec:nirintf}). The errors for the parameters are obtained from the diagonal elements of the covariance matrix of the fit (i. e. the parameters and errors are not marginalized as done for the geometric model in Sect.~\ref{sec:res:ellipse} for computational reasons). The parameters indicate a warm component with a relatively shallow brightness distribution, contributing the majority of the mid-IR emission. The hot component dominates the near-IR emission and gives rise to the $3-5\,\micron$ bump commonly seen in type 1 AGN \citep[e.g.][]{Ede86,Mor09}. Its brightness distribution is very steep, much steeper than for the warm component. Consistent with the ellipse fitting in Sect.~\ref{sec:res:ellipse}, the major axis of the warm component points toward PA $-56^\circ$. We also show results for an alternative model where we fixed the temperature of the hot component to $T_\mathrm{max}^\mathrm{hot} = 1400$\,K, which corresponds to the color temperature in the $J$, $H$, and $K$-band for an accretion disk contribution fraction of 0.24 \citep{Kis07}. The fit is worse than for the best-fit model, but given the small $\chi^2_r$ it still is a credible alternative. In any case comparing the parameters of both models shows the range of uncertainty in the parameters.

An interesting aspect of the models is the relatively low maximum temperature $T_\mathrm{hot}=921\,$K for the hot-dust component, despite the high color temperature. However, it has to be emphasized again that there are model fits at higher temperature that have equally acceptable $\chi^2_r$ values, so that a temperature in the range of 1000$-$1200\,K is well possible. Nevertheless this is lower than what has typically be assumed as the sublimation temperature of 1500$-$1900\,K for graphite dust grains, which supposedly dominate the dust composition closest to the AGN \citep[e.g][]{Bar87,Phi89,Kis07,Kis09b}. The modeled inner radius $R_\mathrm{in}=0.146$\,pc, however, agrees well with the interferometric $K$-band size of 0.16\,pc, given the model and observational uncertainties, respectively. It is, therefore, possible that the maximum temperature in the dust distribution of NGC~3783 is indeed lower than the canonical sublimation temperature. We want to point out that the $K$-band reverberation mapping lag is a factor of $\sim2.5$ shorter than implied by the modeled and near-IR interferometric sizes. This does not necessarily contradict the model parameters, however, as already mentioned in Sect.~\ref{sec:nirintf}: the small reverberation lag is probably a combination of the fact that reverberation techniques are sensitive to the fastest reaction of the variability signal and an inclination-induced projection effect as suggested by the potential ellipticity of the near-IR source.

\begin{figure}
\begin{center}
\epsscale{1.2}
\plotone{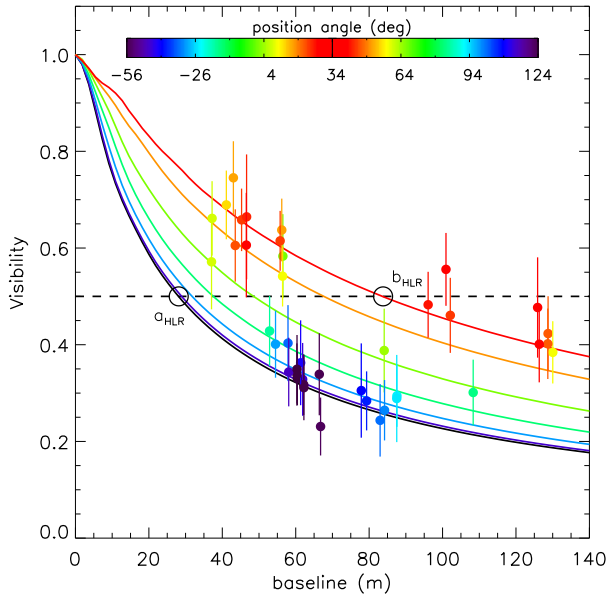}
\caption{12.5\,$\micron$ (restframe) visibilities plotted against the projected baseline length. The position angles for each data point is color coded from dark-blue along the modeled major axis at $-56^\circ$ to red along the minor axis in PA 34$^\circ$. Overplotted are PA-dependent visibilities for the best-fitting model (see Table~\ref{tab:mod:fit}) in steps of $\Delta$PA$=15^\circ$ with color coding according to the respective PA. The dashed line marks the 0.5-visibility level from which the half-light radius of the major axis $a_\mathrm{HLR}$ and the minor axis $b_\mathrm{HLR}$ have been inferred (see text).}\label{fig:res:visbl}
\end{center}
\end{figure}
 
The normalization factors $f_0$ for the hot and warm component can be interpreted as a surface emissivity. For both component we obtain sub-unity values, meaning that the combination of (projected) emitting surface and optical depth is $<$1. If it were the result of a patchy projected emitting surface, only about 10$-$30\% of the projected surface would contribute to the observed emission. Although possible, it seems difficult to reconcile this scenario with the high covering factor of the dust (see Sect.~\ref{sec:cov}). In this regard it is interesting to note that if we assume that most of the projected surface is covered with IR-emitting dust, the emissivity is consistent with mid-IR optically-thin emission. Similar emissivities of about 0.3 have been found in clumpy torus models where the emission is dominated by the optically-thin surface layer of an otherwise optically-thick cloud \citep[``fluffy cloud model'';][see their Fig.~3]{Hon06,Sch08,Hon10b}. In this picture, the radiative transfer effects in a clumpy medium could not be the only effect for observing a shallow silicate emission feature (because of $\tau_\mathrm{IR}<1$), but a non-ISM, graphite-dominated dust composition has to contribute to the feature's weak appearance.

\section{Discussion}\label{sec:disc}

\subsection{Half-light radius and intrinsic elongation}

The fitting result described in the previous section makes use of all 41 data sets and is not restricted to a certain baseline length as in the geometrical model for the 61\,m baseline. Hence, the object geometry is less dependent on the probed spatial scales and the model than the analysis in Sect.~\ref{sec:res:ellipse}. Therefore it is important to realize that the inferred axis ratio $s=a/b=3.0$ reflects the intrinsic geometry more realistically than the axes ratio of the $\sim$1.4 derived from the geometric modeling of the 61\,m data (see Sect.~\ref{sec:res:ellipse} and Figs.~\ref{fig:res:allsizes} \& \ref{fig:midi_inference}). This is well illustrated in Fig.~\ref{fig:res:visbl}. Here we show the observed $12.5\,\micron$ visibilities for all 41 data sets and compare them to the best-fitting model (see Table~\ref{tab:mod:fit}). For the minor axis, the 12.5\,$\micron$ visibility drops to 0.5 at about a baseline length of $\sim$84\,m (red data points). Along the major axis (blue data points), even the shortest baselines at about 50\,m show visibility levels lower than 0.5, meaning that the intrinsic axis ratio must be $>2$. The model predicts that the half-visibility level is reached already at $28.1$\,m, leading to the axis ratio as listed in Table~\ref{tab:mod:fit}. Using the model we can infer a half-light radius at $12.5\,\micron$, despite the low visibility levels along the major axis, of $(20.0\pm3.0)\,\mathrm{mas}\,\times\,(6.7\pm1.0)\,\mathrm{mas}$, corresponding to $(4.23\pm0.63)\,\mathrm{pc}\,\times\,(1.42\pm0.21)\,\mathrm{pc}$ or $(69.4\pm10.8)\,r_\mathrm{in}\,\times\,(23.3\pm3.5)\,r_\mathrm{in}$.

We also tested what kind of effect we may expect on single-telescope mid-IR images of NGC~3783. For that we synthesized a telescope PSF using a 2D Airy function with a core width of $0\farcs35$ as typically measured for calibrator stars in VLT/VISIR images at 12.5\,$\micron$ \citep{Hon10a,Reu10}. This PSF is used as a convolution core for a 12.5\,$\micron$ model image from our best-fit model parameters. We simulated ``observed'' model images, binned to the $0\farcs075$/pixel detector resolution of VISIR, and assumed a detection with SNR=100 of the source and SNR=1000 of a bright PSF reference star. Under these conditions we subtracted the scaled reference star from the ``observed'' model image and measured how much flux is left over from the extended structure. We found that there should not be any detectable extended flux for NGC~3783 based on our best-fit model down to $<$1\%. This is consistent with an analysis of several ground-based mid-IR images by \citet{Asm13}, which did not find any extended structure around the nucleus of NGC~3783. However, if we put NGC~3783 in place of the object with the highest intrinsic spatial resolution in the mid-IR, NGC~1068, ($D$=14.4\,Mpc, $r_\mathrm{in}\approx0.1$\,pc), we would expect that about $10-12$\% of the observed flux could be detected as extended emission in the polar region, with the direction of elongation well recovered. A comparison to Circinus ($D$=4.2\,Mpc, $r_\mathrm{in}\approx0.025$\,pc) yielded similar results ($6-8$\% extended flux). Interestingly, both Circinus and NGC~1068 show much stronger contribution of mid-IR emission from the polar region in single-telescope images than expected from the NGC~3783 model \citep[e.g.][]{Pac05,Mas06}, which may indicate a different intrinsic distribution of the polar dust and/or a viewing-angle effect since both objects are type 2s while NGC~3783 is a type 1 AGN.

\subsection{The covering factors of the hot and warm components}\label{sec:cov}

The two-component model spatially and spectrally disentangles the hot and warm component. As evident from the SED, the hot component occupies primarily the near-IR wavelength region and is responsible for the $3-5\,\micron$ bump while the warm component causes the mid-IR bump. Using the model for decomposing the emission into a hot and warm component, we can determine the observed covering factors of both components as the ratio of the total luminosity in each component to the total luminosity of the UV/optical BBB emission $L_\mathrm{BBB}$. We want to note that these observed covering factors are based on observations along a single line of sight and, therefore, assume isotropy of both the BBB and IR emission. They do not take into account possible anisotropy/optical depth effects of the IR emission, which may be preferentially emitted vertically (type 1 AGN). As a consequence, the intrinsic covering factors in this type 1 AGN are probably smaller than the observed values. The actual degree of anisotropy in the mid-IR is still a matter of debate \citep[e.g.][]{Buc06,Gan09,Hon11a,Asm11}.

We estimate the luminosity of the big-blue bump in NGC~3783 by combining the UV/optical observations compiled in \citet{Pri10} and a ground-based nuclear 4000$-$8000\,\AA~ spectrum \citep[][scaled to the photometric data]{Jon09}, and scaling an AGN BBB template to the extinction-corrected data \citep[multi-power-law template as shown in the appendix of][]{Hon10b}. Assuming that the optical/UV emission in NGC~3783 follows such a template, we integrate the scaled template and obtain $L_\mathrm{BBB} = 1.0 \times 10^{44}\,\mathrm{erg/s}$. This is notably different from the integration of the data itself in \citet[][$3.9\times10^{43}\,$erg/s]{Pri10} where the intention was to quantify the directly observed BBB luminosity. Indeed, NGC~3783 shows signs of extinction toward the shortest wavelengths, and the template fitting implies a line-of-sight optical depth of $\tau_V=0.3$. We estimate the systematic uncertainty of our $L_\mathrm{BBB}$ determination as about a factor of two, primarily due to AGN variability in this source \citep{Onk02}.

Decomposing the IR SED, we obtain $L_\mathrm{hot} = 4.2 \times 10^{43}$\,erg/s and $L_\mathrm{warm} = 9.2 \times 10^{43}$\,erg/s for the hot and warm component, respectively. The total IR luminosity is, therefore, $L_\mathrm{IR} = 1.3 \times 10^{44}$\,erg/s, leading to a total IR covering factor of $C_\mathrm{IR} = L_\mathrm{IR}/L_\mathrm{BBB} = 1.3^{+1.3}_{-0.7}$ (errors based on the optical variability). For the hot and warm components we determine covering factors of $C_\mathrm{hot} = 0.42^{+0.42}_{-0.21}$ and $C_\mathrm{warm} = 0.92^{+0.92}_{-0.46}$ and a ratio $C_\mathrm{warm}/C_\mathrm{hot} = 2.2$,  meaning that the warm dust covers about twice as much solid angle around the AGN as the hot dust. Yet, the hot dust covering factor is still significant and larger than the typical value of $\sim7$\% derived from single-wavelength $K$-band data of a sample of type 1 AGN \citep{Lan11}. 

\subsection{Origin of mid-IR polar emission and the geometry of the dust distribution}

The parsec-scale polar-oriented axis ratio of the mid-IR emission of about 3:1 is the strongest such elongation detected so far. It has to be pointed out again that NGC~3783 is a type 1 AGN with multiple clearly-detected broad lines in the optical as well as low polarization consistent with an equatorial scatterer \citep{Smi04}. Additionally the [\ion{O}{3}] emission appears as only marginally extended \citep{Sch03}. Together with the IR SED and the low X-ray column density this suggests that the type 1 classification is due to a relatively low inclination of the equatorial obscurer (i.e. torus) in NGC~3783 as opposed to a ``lucky strike'' through a hole in a clumpy torus that is seen more edge-on. Therefore the observed strong elongation points toward dust that is significantly elevated from the equatorial plane of the AGN.

We have argued in \citet{Hon12} for the type 2 AGN NGC~424 that the intrinsic polar elongation with axis ratio of 2.2:1 poses a challenge to the standard interpretation of the mid-IR emission with torus models. Indeed, the only models that showed a qualitatively consistent geometry with the observations were the hydrodynamic simulations of \citet[][their Fig.~8]{Sch09}. However, these models failed to reproduce the observed SED. NGC~3783 has a relatively red mid-IR SED \citep[see spectral slopes in ][for a comparison to a small sample of other type 1s]{Hon10a,Kis11b} that might be qualitatively consistent with the model shown in \citet[their Fig.~9]{Sch09}. On the other hand, for more face-on inclinations as in NGC~3783, the model does not reproduce the strong polar elongation, aside from the lack of a near-IR emission bump at around 3--5\,$\micron$\,\,\footnote{It has been speculated that the near-IR bump may be due to the fact that silicates and graphites sublimate at different temperatures and that small grains are cooling less efficiently, leading to a sublimation zone rather than a sublimation radius \citep[e.g.][]{Kis07,Mor12}. The \citet{Sch09} models, however, do account for this effect by separating the grains \citep[similar to the clumpy models in][]{Sch08}, yet no 3--5\,$\micron$ bump is seen.}. Therefore we conclude that the strong pc-scale polar elongation seen in NGC~3783 (as well as some other AGN) is not accounted for in current torus models and challenges the paradigm that the IR emission is dominated by the torus.

\begin{figure}
\begin{center}
\epsscale{1.2}
\plotone{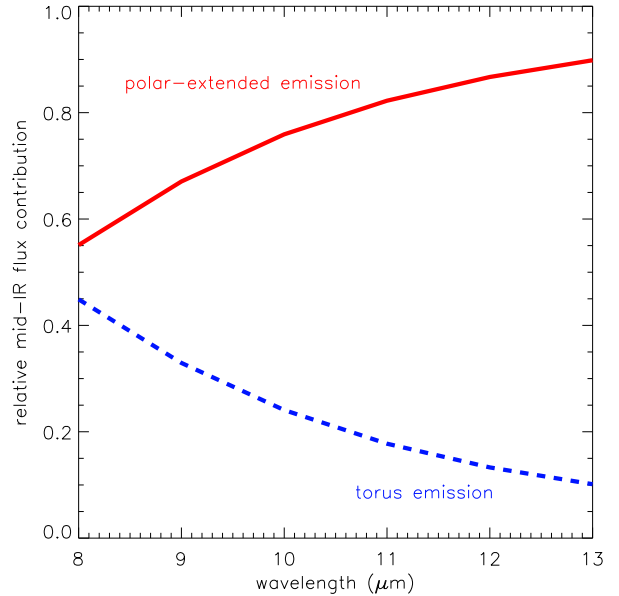}
\caption{Relative contribution of the two model components to the mid-IR emission in NGC~3783 in the $8-13\,\micron$ range. The red-solid line represents the polar-extended component while the blue-dashed line shows the fraction of flux from the torus. These contributions have been calculated based on the best-fitting model from Sect.~\ref{sec:model}.}\label{fig:contrib}
\end{center}
\end{figure}

In order to be more quantitative about how much of the mid-IR flux might still be emitted by the torus, we use our best-fit model from Sect.~\ref{sec:model} and compare the polar-extended flux to the total mid-IR emission. The remaining flux may be associated with the torus. In Fig.~\ref{fig:contrib} we show the relative fraction of polar mid-IR emission and torus emission in the $8-13\,\micron$ range. About 60-90\% of the emission arises from the polar region with a strong dependence on wavelength. This is tied to the overall increase in size toward the polar region with wavelength.

In the previous section we discussed the covering factors of the hot and warm component in NGC~3783 and compared them to the low values in the $K$-band. One possibility to unite the low $K$-band covering factors, the $\sim$30\% $C_\mathrm{hot}$ for NGC~3783, and the factor of 2 larger $C_\mathrm{warm}$ is significant flaring in the dust distribution around the AGN. In this scenario the total surface of dust clouds directly exposed to the AGN grows rapidly with distance (see also Fig.~9 in \citet{Hon12} and Fig. 4 in \citet{Kis13}). Accordingly, the dust has to be elevated significantly. Turbulence seems to be not a very good driver since most hydrodynamic simulations show comparably flat distributions \citep[e.g.][]{Wad09}. In fact, the main reasons why the models by \citet{Sch09} show some polar elongation are the initial conditions which assume that the dust starts out in a more or less spherical distribution on scales of 10s of parsecs (M. Schartmann, private communication). Recently, we argued that radiation pressure on dust both from UV/optical and IR photons may drive an optically-thin dusty wind from the inner region of the torus \citep{Hon12,Kis13}. This wind is not currently accounted for in radiative transfer models that are used to interpret the IR emission of AGN. A notable feature of the torus+wind model are two regions that differ not only in their locations (equatorial torus versus polar wind) but also in their optical depths. This is reflected by the two distinct spectral bumps in the IR SED where the torus would be responsible for the emission peaking in the 3--5\,$\micron$ bump, while the wind mainly contributes to the mid-IR. The spatial distinction is suggested by comparison of the near-IR and mid-IR interferometry as discussed in Sect.~\ref{sec:nirintf}.

We want to emphasize that the observations do not require that the polar cone is filled with dust. The elongation that we see is a projection of the 3D geometry onto the plane of the sky. This means that it is well possible that the polar mid-IR emission is ``hollow'', corresponding to typical wind scenarios \citep[e.g.][]{Elv00,Kea12,Rot12}. In fact, the polar-elongated mid-IR emission in Circinus is brightest close to the edge of the ionization cone \citep{Tri12}. This would help to reconcile polar-extended dust with type 1 AGN being observed without significant optical extinction.

In the context of the wind scenario, it may be interesting to note that NGC~3783 is one of only three type 1 AGN with a (weak) water maser detection (J. Braatz, private communication\footnote{see data at http://www.gb.nrao.edu/$\sim$jbraatz/masergifs/ngc3783.gif}). Although the data are not sufficient to determine the kinematic structure of the maser-emitting clouds in NGC~3783, the other two type 1 AGN with water masers \citep[NGC~4051 and NGC4151;][]{Hag03,Bra04} show outflow characteristics. If the maser emission in NGC~3783 also originated from the outflow region, it supports the idea of outflowing clouds with significant column, potentially bearing dust, in about the polar direction that can emit significant IR emission.

Finally, we want to add that the sub-unity emissivities that favor graphite grains (see Sect.~\ref{sec:model}) are qualitatively consistent with our expectations for the dusty wind. In \citet{Hon12} we discuss that the most likely wind-launching region is the innermost, hottest region of the torus where both the radiation from the AGN as well as from the IR emission of the dust itself has the potential to uplift dust grains via radiation pressure \citep{Kro07}. It is, therefore, conceivable that the dust composition in the wind reflects the dust composition in the hottest region of the torus. Observational evidence strongly suggest that large graphite grains dominate this hot sublimation zone \citep[e.g.][]{Kis07,Kis09b,Mor09,Kis11a,Mor12}.

\section{Summary and conclusions}\label{sec:summary}

We present mid-IR interferometric observations of the type 1 AGN in NGC~3783. The data densely cover the $uv$-plane, which allows us to derive sizes and elongation for the mid-IR emitting source in the wavelength range between $8-13$\,$\micron$. NGC~3783 belongs to a growing sample of objects that have been spatially resolved on (sub-)parsec scale in both the near- and mid-IR, and we made use of the $2.2\,\micron$ data from \citet{Wei12} to complement the mid-IR data. Moreover we compiled a high-angular resolution IR SED as the total flux reference as well as to test the energetics of dust heating and re-emission. Our main results are summarized as follows:

\begin{itemize}
\item The mid-IR emission shows strong elongation and is well resolved with visibilities in the range of 0.2 to 0.75. 
\item Our data can be used to reconstruct sizes for a fixed baseline length of 61\,m. Using a Bayesian approach to model the data, we obtain sizes for the semi-major of $\derr{1.34}{0.06}{0.05}$\,pc at $8.6\,\micron$ and $\derr{2.55}{0.10}{0.08}$\,pc at $12.4\,\micron$, showing a strong wavelength dependence. Qualitatively an increase of size with wavelength is expected for dust that is centrally heated, but the quantitative change of size does also depend on the dust distribution. Indeed, a better reference are sizes scaled for the inner radius of the dust distribution ($r_\mathrm{in} \approx 0.061$\,pc based on near-IR reverberation), resulting in sizes of $20-70\,r_\mathrm{in}$, depending on position angle. This is consistent with \citet{Kis11b} and can be interpreted as a very extended dust distribution compared to other type 1 AGN.
\item As discussed in literature \citep{Kis11b,Hon12}, the half-light radius is a very good size reference for interferometric data since it does not depend on baseline configuration and intrinsic resolution. Owing to the low visibilities of $<0.5$ along the major axis, we are not able to determine a half-light radius directly from the data. However, using a simple power-law model for the wavelength-, baseline-, and position-angle dependence of the visibilities, we infer a half light radius at $12.5\,\micron$ of $(4.23\pm0.63)\,\mathrm{pc}\,\times\,(1.42\pm0.21)\,\mathrm{pc}$ or $(69.4\pm10.8)\,r_\mathrm{in}\,\times\,(23.3\pm3.5)\,r_\mathrm{in}$. Accordingly, the intrinsic axis ratio of the mid-IR emission source is $3:1$
\item The major axis of the mid-IR emission points toward position angle $-52^{+2}_{-3}\,^\circ$, derived from the geometric modeling of the 61\,m data, which is consistent with the PA of $-56^\circ$ obtained by model fitting in Sect.~\ref{sec:model}. This orientation is very close to the polar axis of NGC~3783 inferred from optical spectro-polarimetry ($\Delta$PA$=-7^\circ$). The narrow-line region seems to be misaligned with this polar axis on larger scales ($\ga$50\,pc; $\Delta$PA about $50^\circ$), but apparently changes its direction on smaller scales ($\la$20\,pc; $\Delta$PA about 10$^\circ$) and aligns closer with the polar axis.
\item Given the visibility levels along major and minor axes and a model decomposition, we conclude that about 60$-$90\% of the unresolved single-telescope mid-IR emission originates from the polar region, with a dependency on wavelength. This is consistent with the other two cases where observations with similarly good $uv$-coverage have been performed to date (Circinus, NGC~424).
\item Although the near-IR interferometry reported by \citet{Wei12} still needs follow-up observations, we find tentative evidence that the orientation of the near-IR is inconsistent with the orientation of the mid-IR emission and potentially even perpendicular to the mid-IR emission ($\Delta$PA$\sim82^\circ$ for the peak of the $K$-band PDF). This suggests that the near-IR and mid-IR emission originate from spatially separated dust distributions.
\item The high-angular resolution SED shows two distinct bumps: the regular mid-IR bump (peaking at about $20\,\micron$) and a $3-5\,\micron$ bump that has been observed in many type 1 AGN. The association of the near-IR interferometry with the $3-5\,\micron$ bump and the mid-IR interferometry with the mid-IR bump suggests that the near-IR and mid-IR emission are not only spatially but also spectrally distinct. In this respect it will be important to characterize the size and elongation in the $L$, $M$, and $N$-bands for a large sample of type 1 AGN with the upcoming \textit{MATISSE} instrument at the VLTI.
\item We simultaneously modeled the SED and interferometry of NGC~3783 using a two-component power law representing the hot (near-IR) and warm (mid-IR) emission bumps respectively. We found a very compact hot source with most of the emitting (=directly AGN-heated) dust being located within few times the inner radius in the mid-plane of the putative torus. On the other hand the mid-IR emission originates from a quite shallow dust distribution that extends over a wide range of distances in polar direction. 
\end{itemize}

Our results further support the idea that a significant portion of the mid-IR emission in AGN is associated with dust in the polar region \citep{Hon12,Kis13,Zha13}. It seems as if standard torus models have problems reconciling SED \textit{and} interferometry at the same time, suggesting that the polar dust is physically disconnected from the torus. This dust may arise from a radiatively-driven dusty wind off the inner hottest region of the torus. As such the use of torus models on unresolved IR photometry might have to be reconsidered, and we strongly encourage testing the effect of physically-motivated wind models on dust around AGN \citep[e.g. as in][]{Dor11,Dor12,Kea12,Rot12}.

\acknowledgments

\begin{footnotesize}
\textbf{Acknowledgements ---} We would like to thank Jim Braatz and Francisco Mueller-Sanchez for providing helpful input, as well as the anonymous referee for helpful comments that strengthened the manuscript. This research is based on observations made with the European Southern Observatory telescopes under programs 075.B--0215, 380.B--0289, 082.B--0330, 083.B--0452, 087.B--0401, 088.B--0044, and 089.B--0036. S.F.H. acknowledges support by Deutsche Forschungsgemeinschaft (DFG) in the framework of a research fellowship (“Auslandsstipendium”). This research has made use of the NASA/IPAC Extragalactic Database (NED) which is operated by the Jet Propulsion Laboratory, California Institute of Technology, under contract with the National Aeronautics and Space Administration. Based, in part, on data obtained from the ESO Science Archive Facility.
\end{footnotesize}

\clearpage
\appendix

\section{NGC 3783 VLTI/MIDI interferometry data}

In Fig.~\ref{fig:app:intf} we present the visibilities (gray) and correlated fluxes (blue) for the 40 individual data sets listed in Table~\ref{tab:obs} along with the date of observation, position angle, baseline length and configuration. The data are sorted for position angle. On top of the data we plot the best-fitting model (green-dashed line) with parameters according to Table~\ref{tab:mod:fit}. The black circles with error bars indicate the binned version of the visibilities that were used in the model fitting process. In some data sets a small residual at the position of the atmospheric ozone feature at $9.6\,\micron$ can be detected, which remained after calibrating the data.

\begin{figure*}[b]
\begin{center}
\epsscale{1.0}
\plotone{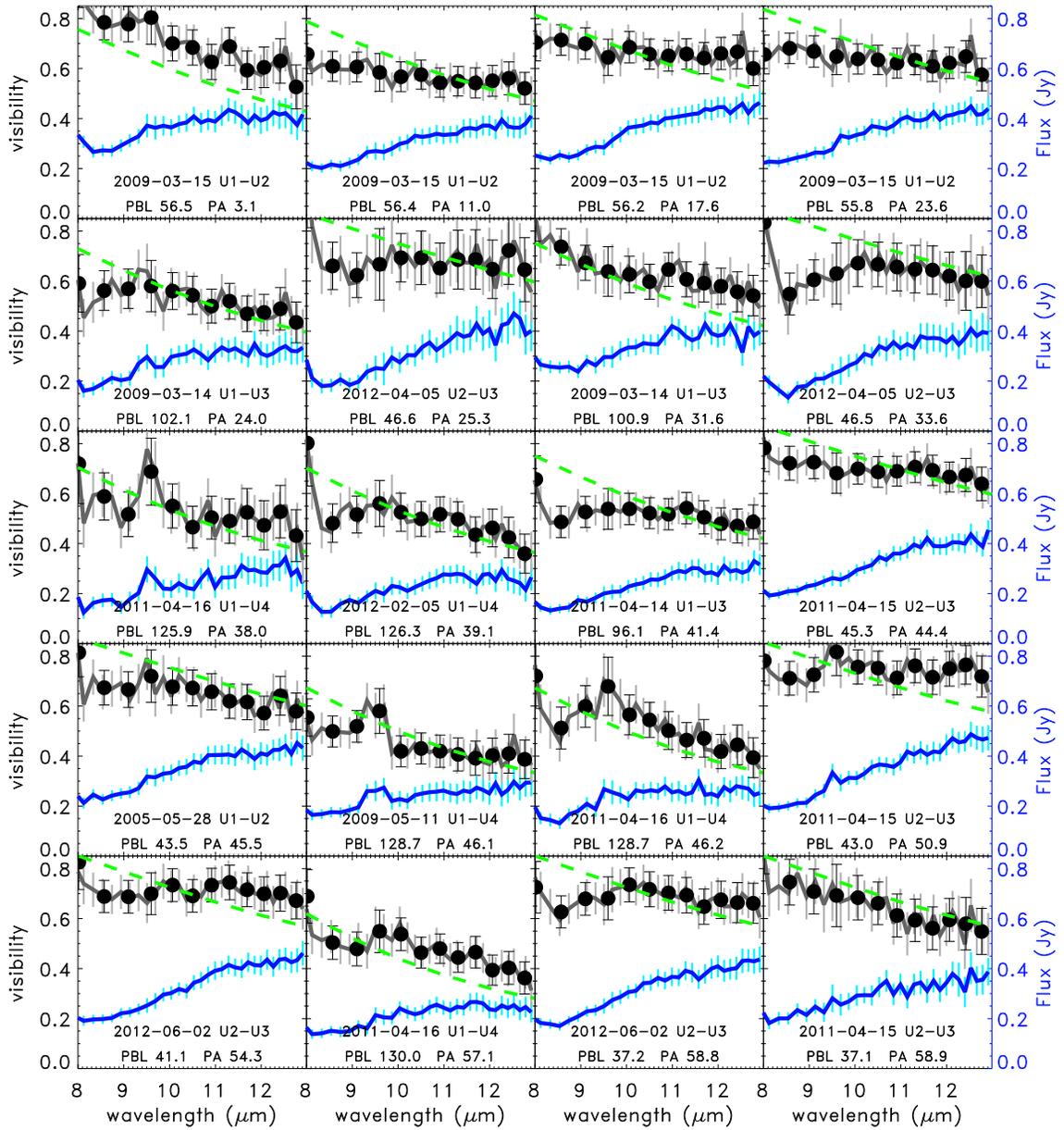}
\caption{MIDI data of NGC 3783 obtained between 2005 and 2012. Blue lines (with error bars) show the correlated fluxes for each baseline configuration that has been observed. Corresponding visibilities are plotted as gray lines (full wavelength resolution of the data) and black circles with error bars (binned data used for model fitting). The green-dashed lines represents the best-fitting model with parameters listed in Table~\ref{tab:mod:fit}.}\label{fig:app:intf}
\end{center}
\end{figure*}

\addtocounter{figure}{-1}
\begin{figure*}
\begin{center}
\epsscale{1.0}
\plotone{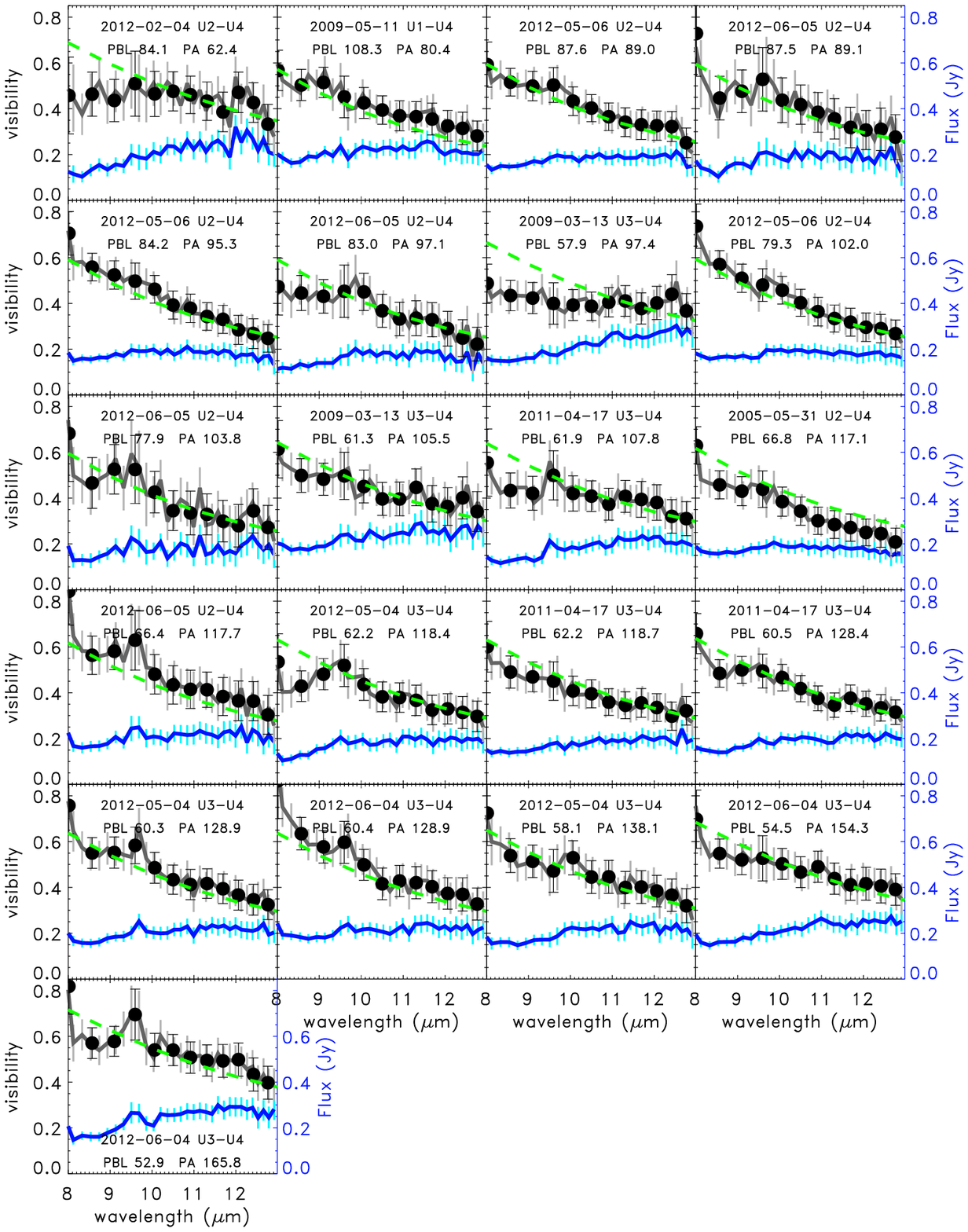}
\caption{continued.}
\end{center}
\end{figure*}

\end{document}